\documentclass[preprint,12pt,onecolumn]{elsarticle}

\usepackage{amssymb}
\usepackage{amsmath,amsfonts}
\usepackage{graphicx}
\usepackage{subfig}
\usepackage{textcomp}
\usepackage{booktabs}
\usepackage{xcolor}
\usepackage{url}
\usepackage{makecell}
\usepackage{booktabs}
\usepackage{siunitx}
\usepackage{array}
\usepackage{tcolorbox}
\usepackage{longtable}
\usepackage{boldline}
\usepackage{makecell}
\usepackage{svg}
\usepackage[utf8]{inputenc}
\usepackage{float}
\usepackage{multirow}
\usepackage{tabularx}
\usepackage{algorithm}
\usepackage[noend]{algpseudocode}
\usepackage[absolute]{textpos}
\usepackage{listings}
\usepackage{ulem}
\usepackage{lineno}
\usepackage{adjustbox}
\usepackage{changepage} 
\usepackage{pgfplots}
\pgfplotsset{compat=1.17}
\usepackage{pgfplotstable}
\usepackage{hyperref}

\journal{Journal of Systems and Software}

\begin{document}

\begin{textblock*}{\paperwidth}(0pt,0pt)
    \vspace{2mm}
    \footnotesize
    \noindent \centering
    \begin{minipage}{0.7\paperwidth}
    \begin{tcolorbox}[left=0mm,right=0mm,boxrule=0.1mm,colback=white!30!white]
        \vspace{-2mm}
        \centering\textit{This is a PREPRINT version accepted to be published in the Journal of Systems and Software (\url{https://doi.org/10.1016/j.jss.2025.112575}). Users are expected to adhere to the terms and conditions set by each copyright holder when reproducing or making these documents available.}
        \vspace{-2mm}
    \end{tcolorbox}
    \end{minipage}
\end{textblock*}

\begin{frontmatter}

\title{A Large-Scale Study on Developer Engagement and Expertise in Configurable Software System Projects}

\author{Karolina M. Milano$^{a}$, Wesley K. G. Assunção$^{b}$, Bruno B. P. Cafeo$^{c}$} 

\address{$^{a}$ Federal Institute of Mato Grosso do Sul, Jardim, Brazil}
\address{$^{b}$ North Carolina State University, Raleigh, USA}
\address{$^{c}$ University of Campinas, Campinas, Brazil}

\begin{abstract}
Modern systems operate in multiple contexts making variability a fundamental aspect of Configurable Software Systems (CSSs). Variability, implemented via pre-processor directives (e.g., \texttt{\#ifdef} blocks) interleaved with other code and spread across files, complicates maintenance and increases error risk. Despite its importance, little is known about how variable code is distributed among developers or whether conventional expertise metrics adequately capture variable code proficiency. This study investigates developers’ engagement with variable versus mandatory code, the concentration of variable code workload, and the effectiveness of expertise metrics in CSS projects. We mined repositories of 25 CSS projects, analyzing 450,255 commits from 9,678 developers. Results show that 59\% of developers never modified variable code, while about 17\% were responsible for developing and maintaining 83\% of it. This indicates a high concentration of variable code expertise among a few developers, suggesting that task assignments should prioritize these specialists. Moreover, conventional expertise metrics performed poorly---achieving only around 55\% precision and 50\% recall in identifying developers engaged with variable code. Our findings highlight an unbalanced distribution of variable code responsibilities and underscore the need to refine expertise metrics to better support task assignments in CSS projects, thereby promoting a more equitable workload distribution.

\end{abstract}


\end{frontmatter}

\section{Introduction}

Due to the pervasive nature of software, modern systems are designed with flexibility as a requirement to operate in a multitude of contexts~\cite{capilla2013systems}. To achieve that, variability is a fundamental aspect of software, enabling the development of \textit{Configurable Software Systems (CSSs)} that can adapt to different market segments or contexts of use~\cite{ berger2020state}. 
Unlike conventional single systems, CSSs are composed of both mandatory and variable features. 
Variable features are usually managed with pre-processor directives (i.e., \texttt{\#ifdef} blocks), which are activated by user-defined configurations~\cite{apel2013book, Michelon2021_GPCE}.  
This approach introduces \textit{variable code}—segments of code that are only included under specific conditions, leading to a scattered codebase where feature-related logic can be dispersed across multiple files. 
Despite the flexibility of pre-processor directives to derive tailored software products, they make the code hard to understand and error-prone to maintain~\cite{medeiros2017discipline}. This requires developers of CSS projects to be experts in variable code. Thus, expertise in variable code plays a crucial role in the maintenance and evolution of CSS projects~\cite{Michelon2023_SANER}.

Developers' expertise is typically derived from developers’ contributions to the codebase and their interactions with specific source files. Expertise metrics are widely used to identify individuals with in-depth knowledge of certain parts of the system~\cite{anvik06,kagdi08}. This knowledge is invaluable for task assignment, risk mitigation, and understanding knowledge distribution within a project~\cite{anvik07}. Traditional expertise metrics, such as Degree of Authorship (DOA)~\cite{fritz2014degree} and Ownership~\cite{bird2011don}, often focus on file-level contributions to determine which developers have the most experience with specific files, namely experts within particular sections of a codebase. These metrics are critical for preventing knowledge silos, which can threaten the sustainability of a project when a few developers hold a disproportionate amount of knowledge about key areas~\cite{avelino2016,ferreira2017}.

Expertise metrics are mainly studied on conventional single software systems, where the codebase is relatively static, differently from CSSs~\cite{michelon2021hybrid}. In single software systems, developers work on well-defined, self-contained components or files, making it easier to identify expertise based on code contributions. However, in the context of CSSs, expertise management faces a unique challenge due to the inherent variability of these systems. 
Then, we hypothesize that, given the scattered and interleaved nature of variable code~\cite{kastner2008granularity,Michelon2023_SANER,medeiros2017discipline}, the effectiveness of expertise metrics may be significantly impacted.
Expertise metrics probably fail to capture a developer’s engagement with features that are conditionally included or spread across various files. Consequently, developers with significant expertise in variable code may be overlooked by these metrics, and developers identified as ``experts'' may lack the necessary experience with the dynamic aspects of CSSs. 
This can lead to several challenges, including suboptimal task assignment, in which developers may struggle with unfamiliar and complex variable code, whereas experts in these areas remains underutilized. This misalignment can lead to inefficiencies, bottlenecks, and increased maintenance costs across various aspects of the development process~\cite{petersen2014reasons,Rejab2014}.

The goal of this study is to explore the engagement of developers with variable code, and whether traditional metrics can adequately capture expertise in the context of CSSs. Thus, we investigate the relationship between expertise metrics and the distribution of developers' workload in CSSs. We aim at understanding to what extent these metrics are suitable for supporting expertise management in CSS projects.

To achieve this goal, we conducted a large-scale mining repository study across 25 CSS projects, analyzing 450,255 commits made by 9,678 developers, to address three research questions (RQs), as follows:

\begin{itemize}
    \item \textit{RQ1 (Developers Engagement):} How are developers engaged with variable and mandatory code segments? Our analysis revealed that approximately 60\% of development effort is concentrated on mandatory code, with the remaining 40\% focused on variable code.
    \item \textit{RQ2 (Concentration of Workload on Variable Code):} How concentrated is the workload on variable code among developers? We found that few (approx. 17\%) developers contribute to 83\% of the variable code, indicating a concentration of expertise in a small group of individuals.
    \item \textit{RQ3 (Expertise Metrics and Variable Code Engagement):} How do expertise metrics perform in comparison to the actual engagement of developers with variable code? Our findings reveal that while expertise metrics have a precision of 55\% in identifying developers involved in variable code, their recall is only 50\%, indicating significant room for improvement.
\end{itemize}

The results of our study emphasize the limitations of conventional expertise metrics when applied to CSSs, particularly in identifying developers engaged with variable code. Our findings evidence that variable code is maintained by a small subset of developers, highlighting an unbalanced distribution of workload. This underscores the need for refining expertise metrics to better capture the complexities associated with variable code. Additionally, we stress the importance of promoting a more equitable distribution of workload by cultivating more expertise in variable code among developers. Addressing these limitations will contribute to more accurate expertise management, improving both the maintainability and long-term sustainability of CSS projects.

This paper makes several contributions. Firstly, it provides the first comprehensive evaluation of expertise metrics in the context of variable code within CSSs, extending our preliminary work~\cite{Milano2024}. Secondly, it presents an empirical study of code expertise and workload distribution in CSS projects, offering insights into the challenges of managing expertise specific to variable code. Thirdly, it identifies key weaknesses in existing expertise metrics when applied to CSS projects, demonstrating that these metrics often fail to capture expertise in variable code due to its scattered and interleaved nature. Lastly, it offers a set of lessons learned regarding workload dynamics, expertise distribution, and the precision of expertise metrics in CSS projects.

This paper is organized as follows: Section~\ref{sec:preliminaries} presents the background and introduces the problem statement. Section~\ref{sec:study} details our study design, with results presented in Section~\ref{sec:results}. Section~\ref{sec:implications} discusses the study's implications, while Section~\ref{sec:relatedwork} reviews the relevant literature. Section~\ref{sec:threats} addresses threats to validity, and Section~\ref{sec:conclusion} concludes the paper, highlighting directions for future work.

\section{Background and Problem Statement}
\label{sec:preliminaries}

For CSSs implemented in the C/C++ programming language, the C/C++ preprocessor's ability to handle boolean expressions constructed using macros proves invaluable for managing variabilities, playing a crucial role in managing variability within the implementation code. The preprocessor acts as a powerful tool for conditional compilation, allowing developers to selectively include or exclude code segments based on macro expressions, particularly using the \texttt{\#ifdef} directive. This enables the creation of modular code that adapts to different configurations and feature sets.

The adaptable granularity of variabilities facilitated by conditional compilation is a key advantage of using the C/C++ preprocessor. This flexibility allows variability of code to permeate throughout the program, enabling developers to control the inclusion or exclusion of code elements at various levels, from individual variables and functions to entire code blocks. This granular control is particularly beneficial for scenarios where variabilities are scattered throughout the codebase, making it challenging to isolate and manage them using other techniques.

To illustrate the concept of preprocessor-based variability management in C/C++ and the idea of variable code, consider the code snippet in Listings~\ref{lst:conditional_compilation}. This illustrative code segment demonstrates the use of the \texttt{\#ifdef} directive to control the inclusion of code related to dynamically managing the parser's stack to avoid overflow in the GCC project\footnote{https://github.com/gcc-mirror/gcc}.

\begin{lstlisting}[language=C, xleftmargin=0.5cm, basicstyle=\scriptsize, label=lst:conditional_compilation, caption={Excerpt of file \texttt{parser-scan.c} (GCC Project)}]
#ifdef yyoverflow
      /* Each stack pointer address is followed by the size of
	 the data in use in that stack, in bytes.  */
    #ifdef YYLSP_NEEDED
        /* This used to be a conditional around just the two extra args,
        but that might be undefined if yyoverflow is a macro.  */
        yyoverflow("parser stack overflow",
         &yyss1, size * sizeof (*yyssp),
         &yyvs1, size * sizeof (*yyvsp),
         &yyls1, size * sizeof (*yylsp),
         &yystacksize);
    #else
        yyoverflow("parser stack overflow",
         &yyss1, size * sizeof (*yyssp),
         &yyvs1, size * sizeof (*yyvsp),
         &yystacksize);
    #endif

    yyss = yyss1; yyvs = yyvs1;
    
    #ifdef YYLSP_NEEDED
        yyls = yyls1;
    #endif    
#else /* Extend the stack our own way.  */    
    if (yystacksize >= YYMAXDEPTH) {
        yyerror("parser stack overflow");
        return 2;
    }
    yystacksize *= 2;
    
    if (yystacksize > YYMAXDEPTH)
        yystacksize = YYMAXDEPTH;
        
    yyss = (short *) alloca (yystacksize * sizeof (*yyssp));
    __yy_memcpy ((char *)yyss, (char *)yyss1, size * sizeof (*yyssp));
    
    yyvs = (YYSTYPE *) alloca (yystacksize * sizeof (*yyvsp));
    __yy_memcpy ((char *)yyvs, (char *)yyvs1, size * sizeof (*yyvsp));
    
    #ifdef YYLSP_NEEDED
        yyls = (YYLTYPE *) alloca (yystacksize * sizeof (*yylsp));
        __yy_memcpy((char *)yyls, (char *)yyls1, size * sizeof (*yylsp));
    #endif
#endif /* no yyoverflow */
\end{lstlisting}

\bigskip

In the provided code, part of the \texttt{parser-scan.c} file has \texttt{\#ifdef} directives used to conditionally include code segments based on the presence of specific configuration constants. This file consists of 2,400 lines of code (LOC) and includes 28 configuration constants (macros), used to include or
exclude code segments related to features of GCC. This code handles parser stack overflow in two ways, depending on whether the \textit{configuraton constant} \texttt{yyoverflow} is defined. If \texttt{yyoverflow} is defined (i.e., true), the parser calls the \texttt{yyoverflow} function, which handles reallocating the stack. If \texttt{YYLSP\_NEEDED} is also defined (for location tracking), an additional argument for the location stack is passed to yyoverflow. If \texttt{yyoverflow} is not defined, the parser manually doubles the stack size using \texttt{alloca} and copies the old stack content into the newly allocated space. It also checks if the stack size exceeds the maximum allowed (\texttt{YYMAXDEPTH}), triggering an error if so. Again, if \texttt{YYLSP\_NEEDED} is defined, location stack handling is included.

Beyond the technical details of the given code, this example also highlights important aspects of collaborative software development. The file, \texttt{parser-scan.c}, has been touched by 16 developers over its evolution, with two experts identified by an expertise metric (DOA~\cite{fritz2014degree}), suggesting they theoretically have more knowledge of the file content. However, the variable code associated with the \texttt{yyoverflow} and \texttt{YYLSP\_NEEDED} configuration constants reveals that: (i) variable code related to these configuration constants have never been touched by any of these experts; and (ii) code segments related to these configuration constants were only modified by three developers out of 16 that modified the file during the evolution.

\bigskip

\noindent\textbf{Uneven Workload Distribution.} The inherent complexity of managing variability in CSSs presents a significant challenge for development teams~\cite{Michelon2021_GPCE}. Features heavily reliant on conditional compilation directives can create a disproportionate burden for assigned developers~\cite{Michelon2023_SANER}. Maintaining these variabilities necessitates navigating intricate logic with a multitude of conditional directives and interdependencies~\cite{cafeo2015relationship}. This complexity significantly increases cognitive load and development time, potentially hindering project progress, delaying timelines, and impacting team morale~\cite{steinmacher2015systematic}. Further compounding this issue is the limited understanding of variability implementation among many developers. The inherent complexity of preprocessed CSSs can create a scenario where few possess a comprehensive grasp of how these variabilities are implemented. Variable code expertise may become concentrated among a subset of developers who specialize in managing specific variabilities, creating inadvertent bottlenecks. This concentration of knowledge introduces an imbalance in the workload of variable code as well as project risk if these key individuals are unavailable, hindering knowledge sharing and collaboration~\cite{avelino2016,ferreira2017}.

\bigskip

\noindent\textbf{Inadequacy of Existing Expertise Metrics.} Although expertise metrics have been proven effective in conventional single software systems for task assignment and risk mitigation, their use in CSSs remains constrained. These metrics, typically file-based, may fail to capture the nuanced skill sets required for effective variability management in CSS projects. Developers deemed ``experts'' by conventional metrics may lack proficiency in handling conditional compilation logic or managing fine-grained variability. The fragmented nature of variabilities can deviate from the modular structure assumed by conventional metrics, leading to misjudgments regarding the distribution of expertise within the team.

\bigskip

\noindent\textbf{Consequences.} These challenges associated with variability management in CSSs have far-reaching consequences:

\begin{itemize}
    \item \textbf{Inefficient Task Assignment:} Misalignment between developer expertise and task assignment leads to suboptimal resource utilization, hindering project progress and increasing the risk of errors.
    
    \item \textbf{Knowledge Silos and Collaboration Barriers:} The concentration of expertise among a limited group of developers impedes knowledge sharing and collaboration of variable code, hindering innovation and problem-solving.
    
    \item \textbf{Project Risks:} The reliance on a few key individuals for variability management creates a single point of failure, increasing the risk of project delays or failures if these individuals are unavailable.
\end{itemize}

\smallskip

\noindent\textbf{Understanding the Challenges.} A nuanced understanding of the challenges in expertise management within CSS projects is crucial for advancing both theoretical and practical aspects of software engineering. By highlighting the limitations of conventional expertise metrics, which may not adequately capture the intricacies of variable code, we enhance our comprehension of how these systems operate and identify significant gaps in current practices. This deeper insight into expertise management has broader implications, impacting project efficiency, knowledge dissemination, task assignment, and system maintainability. This foundational knowledge sets the stage for future research and practice improvements, ultimately contributing to more effective management and utilization of expertise within CSSs.

\section{Study Design}\label{sec:study}

The design of our study is based on the Goal/Question/Metric (GQM) approach~\cite{basili1994goal,Basili2002}, as described below.

\subsection{Study goal} \label{sec:goal}

The goal of this study is \textit{to establish an empirical foundation for understanding the engagement and workload of developers, alongside expertise assessment, regarding variable code in CSS projects}. Through an in-depth analysis of the evolutionary history of 25 open-source systems, we explore how variable code engagement evolves within development teams and assess whether established file-based metrics accurately reflect this expertise distribution. By investigating this understudied area, we aim to provide insights for informed decision-making and improved knowledge management in CSS projects.

\subsection{Research Questions} \label{sec:rq}

To accomplish the goal of this study, we address three fundamental research questions, which are described below with the corresponding metrics used to answer them.

\bigskip

\noindent\textit{RQ1: How are developers engaged with variable and mandatory code segments?}

\smallskip

\noindent\underline{Rationale:} Understanding how developers engage with variable and mandatory code is important for optimizing team collaboration, identifying expertise areas, and ensuring balanced involvement in key parts of a system. By examining developer interactions with both variable and mandatory code, we can uncover patterns of expertise and collaboration. This insight is vital for enhancing code quality, managing the inherent complexity of CSSs, and ensuring that developer engagement aligns with technical priorities. By understanding how developers interact with different types of code, teams can better focus their efforts, streamline collaboration, and make more informed decisions that lead to improved system maintainability and overall project success.

\noindent\underline{Metrics:} To answer RQ1, we conducted a systematic and longitudinal approach. Based on their modifications in commits, we classified developers in three categories: (i) \textit{Generalists}: developers exclusively engaged in modifying mandatory code; (ii) \textit{Specialists}: developers exclusively dedicated to altering variable code; and (iii) \textit{Mixed}: developers contributing to both variable and mandatory code. Then, we applied Pearson correlation~\cite{freedman2007} to investigate whether the proportion of mandatory code relative to variable code can explain the developer distribution.
Details are provided in Section~\ref{sec:eval}.

\bigskip

\noindent\textit{RQ2: How concentrated is the workload about variable code among developers?}

\smallskip

\noindent\underline{Rationale:} Investigating the concentration of developer engagement with variable code is crucial for identifying potential overreliance on a limited group of developers in CSSs projects. This analysis seeks to uncover whether a small number of developers primarily handle variable code, which can lead to workload imbalances, knowledge silos, and decreased team efficiency. Literature suggests strategies such as implementing team code ownership/authorship, adopting overlapping pair rotation, fostering open communication, and monitoring workload distribution to address these issues~\cite{Sedano2016}. Understanding these dynamics is vital for promoting equitable participation, enhancing collaboration, and reducing risks associated with concentrating critical knowledge and responsibilities on a few individuals, ultimately leading to more sustainable and resilient development practices.

\noindent\underline{Metrics:} 
For RQ2, we explore the concentration and distribution of variable code development for mixed developers and specialists (classified in RQ1).
To analyze the equality or inequality in the distribution of work among developers, we performed a data aggregation approach by concentration statistics~\cite{gastwirth1972estimation}. We applied Lorenz statistical method~\cite{gastwirth1972estimation}to examine inequalities within developers. 
To compare Lorenz concentrations, we computed the Gini coefficient~\cite{dorfman1979formula}. Details of these metrics are provided in Section~\ref{sec:eval}. 

\bigskip

\noindent\textit{RQ3: How do expertise metrics perform in comparison to the actual engagement of developers with variable code?}

\smallskip

\noindent\underline{Rationale:} Expertise metrics are commonly used to identify key contributors in software projects, often labeling developers as experts based on their historical contributions to specific files~\cite{avelino2019measuring}. However, it is pivotal to determine whether these designated experts are consistently involved in developing and maintaining variable code, which is typically more complex and prone to change within CSS projects. Understanding this alignment is essential for assessing the reliability of expertise metrics and ensuring they accurately reflect actual developer engagement with critical code segments. A positive outcome reinforces the credibility of expertise metrics, supporting their use in guiding expertise management in CSS projects. Conversely, a lack of engagement from designated experts could expose gaps in expertise allocation and highlight the need for more refined methods to better align expertise in CSSs, ultimately improving project management and reducing technical risk.

\noindent\underline{Metrics:}  
We compute two code expertise metrics, namely Degree of Authorship~\cite{fritz2014degree} and Ownership~\cite{bird2011don}, detailed in Section~\ref{sec:expert_metrics}. These metrics are the basis for investigating whether developers designated as file experts are involved in modifying the variable code of their files. To quantify the alignment, we computed Precision and Recall~\cite{Ting2010}.

\subsection{Subject Systems} \label{sec:systems}

This study conducts an analysis of real-world, open-source CSS projects. Our initial dataset comprised 65 preprocessor-based systems, all implemented in the C/C++ programming language. These systems are publicly hosted on GitHub and have been the subject of extensive examination in prior studies~\cite{kikas2016using, abal2018variability, liebig2010analysis, medeiros2013investigating, medeiros2017discipline, rodrigues2016assessing}. To refine the scope of our analysis, we manually narrow it down from 65 to 25 systems that meet inclusion criteria, specifically those with (i) more than 30 developers, and (ii) featuring a minimum of 50 configuration constants (i.e., at least 50 different conditions used in \texttt{\#ifdef} directives and that act as guards for the code that is enclosed between an opening \texttt{\#ifdef} and its closing complement).

Table~\ref{tab:project-stats} presents an overview of our 25 subject systems, organized into three thematic blocks. Under “Repository Info” we list commits (ranging from 1,987 in Libexpat to 133,378 in GCC) and developers (33 in Lighttpd1.4 up to 1,739 in GCC). The “Structural Properties” block shows file counts (130-83,245) and configuration constants (214–28,510), while the “Variability Info” block reports total lines of code (6,128–1,431,731) and the proportion of mandatory vs.\ variable code (from 31.13 \% mandatory in Libxml2 to 98.78 \% in Totem, and the complementary 1.22 \%–68.87 \% variable).

\begin{table*}[t]
    \scriptsize
    \caption{Project Characteristics and Descriptive Statistics}
    \label{tab:project-stats}
    \centering
    \addtolength{\tabcolsep}{-2pt}
    \begin{tabular}{l|
                    r  r |  % Repository Info
                    r  r   % Structural Properties
                    | r  r  r} % Variability Info
      \toprule
      & \multicolumn{2}{c|}{\textbf{Repository Info}} 
      & \multicolumn{2}{c|}{\textbf{Structural Properties}} 
      & \multicolumn{3}{c}{\textbf{LOC Info}} \\
      \cmidrule(lr){2-3} \cmidrule(lr){4-5} \cmidrule(lr){6-8}
      \textbf{Project} 
        & \textbf{\# Commits} 
        & \textbf{\# Devs.} 
        & \textbf{\# Files} 
        & \textbf{\# C.C.} 
        & \textbf{LOC}  
        & \textbf{\% Mand.} 
        & \textbf{\% Var.} \\
      \midrule
      libxml2         &   4522 &  223 &   261 &  2590 &  241609 & 31.13\,\% & 68.87\,\% \\
      curl            &  16674 &  824 &  1315 &  3119 &  160491 & 32.61\,\% & 67.39\,\% \\
      mapserver       &   8787 &  126 &   874 &  1596 &  135137 & 54.86\,\% & 45.14\,\% \\
      gcc             & 133378 & 1739 & 83245 & 28510 &  818058 & 55.44\,\% & 44.56\,\% \\
      openvpn         &   2829 &  138 &   525 &   994 &   96819 & 56.81\,\% & 43.19\,\% \\
      marlin          &  15597 & 1082 &  7780 & 12429 &  150410 & 63.57\,\% & 36.43\,\% \\
      libssh-mirror   &   4644 &  126 &   349 &   402 &   77275 & 63.84\,\% & 36.16\,\% \\
      openssl         &  22477 &  761 &  3507 &  3550 &  423560 & 66.73\,\% & 33.27\,\% \\
      emacs           &  42073 &  398 &  1510 &  5647 &  484489 & 69.29\,\% & 30.71\,\% \\
      ossec-hids      &   2632 &   88 &   666 &   823 &   67381 & 70.41\,\% & 29.59\,\% \\
      gnuplot         &   7015 &  126 &   333 &  1447 &  123488 & 74.54\,\% & 25.46\,\% \\
      retroarch       &  48950 &  536 &  7926 & 12053 &  714357 & 74.72\,\% & 25.28\,\% \\
      lighttpd1.4     &   4027 &   33 &   282 &  1154 &   84854 & 76.96\,\% & 23.04\,\% \\
      uwsgi           &   5189 &  300 &   301 &   415 &   76039 & 77.77\,\% & 22.23\,\% \\
      glibc           &  23219 &  325 & 20702 &  7042 &  296791 & 78.30\,\% & 21.70\,\% \\
      libexpat        &   1987 &   59 &   130 &   214 &   27348 & 80.63\,\% & 19.37\,\% \\
      mongo           &  52545 &  686 & 70412 & 27905 & 1431731 & 81.94\,\% & 18.06\,\% \\
      busybox         &  14844 &  378 &  1406 &  3344 &  139420 & 89.79\,\% & 10.21\,\% \\
      xorg-xserver    &  14826 &  617 &  2861 &  3318 &  348293 & 91.22\,\% & 8.78\,\% \\
      hexchat         &   2089 &  165 &   358 &   411 &   51353 & 92.84\,\% & 7.16\,\% \\
      collectd        &   6708 &  486 &   645 &   847 &   75447 & 93.22\,\% & 6.78\,\% \\
      dia             &   3885 &   76 &   842 &   475 &   42340 & 95.62\,\% & 4.38\,\% \\
      irssi           &   4287 &   99 &   504 &   407 &   10529 & 96.68\,\% & 3.32\,\% \\
      libsoup         &   2353 &  159 &   478 &   321 &   44864 & 96.97\,\% & 3.03\,\% \\
      totem           &   4718 &  128 &   459 &   344 &    6128 & 98.78\,\% & 1.22\,\% \\
      \midrule
      \textbf{Minimum}    &   1987 &   33 &   130 &   214 &    6128 & 31.13\,\% & 1.22\,\% \\
      \textbf{Maximum}    & 133378 & 1739 & 83245 & 28510 & 1431731 & 98.78\,\% & 68.87\,\% \\
      \textbf{Average}    &  1949.4 &1901.1& 1922.9& 1913.1&   2571.2 & 61.13\,\% & 38.87\,\% \\
      \textbf{Median}     &   6708 &  223 &   666 &  1447 &  123488 & 76.96\,\% & 23.04\,\% \\
      \textbf{Mode}       &     -- &  126 &    -- &    -- &      -- & -- & -- \\
      \textbf{Std.\ Dev.} & 28102.9&397.5 &21158.9& 7809.3&  326363.1 & 0.19\,\% & 0.19\,\% \\
      \textbf{Skewness}   &    3.22 & 1.95 &  3.13 &   2.43 &    2.46 & --0.78  & 0.78 \\
      \textbf{Kurtosis} & 12.07 & 4.56 & 9.00 & 5.42 & 6.78 & 0.30 & 0.30 \\
      \bottomrule
    \end{tabular}
\end{table*}

\begin{figure*}[t]
  \centering
  \subfloat[\# of Developers (log scale)]{%
    \includegraphics[width=0.30\textwidth]{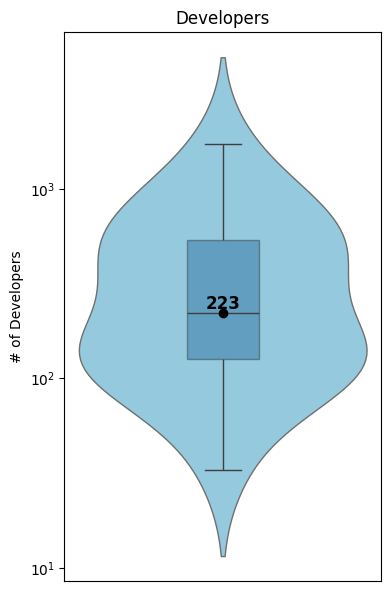}
  }
  \hfill
  \subfloat[Configuration Constants (log scale)]{%
    \includegraphics[width=0.30\textwidth]{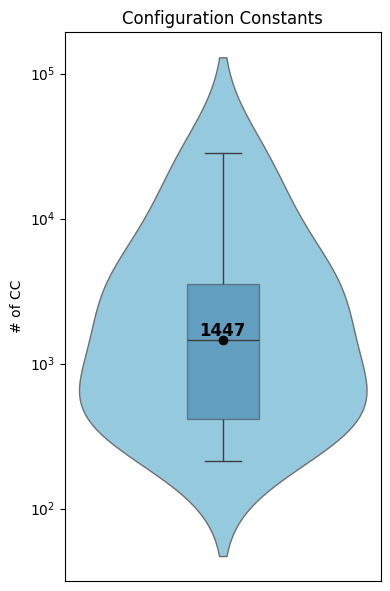}
  }
  \hfill
  \subfloat[\% Variable Code]{%
    \includegraphics[width=0.30\textwidth]{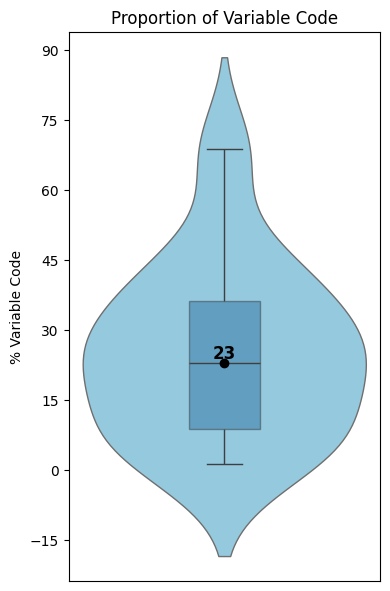}
  }
  \caption{Distribution of repository, structural and variability-related characteristics across subject systems. Medians are annotated.}
  \label{fig:project-violin-plots}
\end{figure*}

To visually reinforce these patterns, in Figure~\ref{fig:project-violin-plots}, we include three boxplots with violin that capture the distribution of developer profiles, configuration constants, and the proportion of variable code across the subject systems.

\subsection{Ground Truth Construction} \label{sub:collection}

We outline the comprehensive construction of the ground truth of our study, which consists of four fundamental steps. These steps are designed to provide us with the necessary data to effectively address the RQs of our study (see Section~\ref{sec:rq}). 
All our source code, data, and supplementary materials can be publicly accessed online~\cite{replicationpackage}.

\bigskip

\noindent\textbf{Development History Extraction.} To initiate the data collection process, we started by cloning and extracting the development history from the aforementioned repositories. This process is facilitated through the use of a custom script developed using Pydriller~\cite{PyDriller}, a Python framework crafted to aid developers in analyzing Git repositories. With this script, we systematically processed each commit, capturing three essential pieces of information: (i) the file path of changed files, (ii) the developer responsible for the change in a commit, and (iii) the type of change made---whether it was an addition, modification, or deletion---in each line of the changed files. We should note that files that are not source code, such as images or documentation, are filtered out from the dataset. By excluding files without source code, we maintain the data's quality and relevance, aligning with best practices in empirical software engineering research. In addition, systematically processing each commit within the repositories, we ensure a representative sample of the development history. Contributions from branches were included only if merged into the main branch to ensure consistency in tracking changes. To handle tangled commits, our line-by-line analysis attributed changes to the responsible developer regardless of commit complexity. This approach ensures that modifications to variable code are accurately associated with developers, even in cases where multiple features are altered in a single commit. This approach allows us to draw meaningful insights and make inferences about the entire population of code changes, developers, and their interactions. 

\bigskip

\noindent\textbf{Information Extraction of Variable Code.} With the development history and code files at our disposal, we proceed to extract variable code information. To achieve this, we enhance the functionality of a Python-based tool, pypreprocessor,\footnote{\url{https://github.com/interpreters/pypreprocessor}} which is a C-style macro preprocessor. Our modified version of pypreprocessor is tailored to identify program elements and lines of code that constitute variable code. In essence, we perform a meticulous line-by-line analysis of the source code to pinpoint segments that contribute to variable code. Files that do not contain any variable code are excluded from our dataset. By performing a line-by-line analysis, we ensure that each part of the source code, whether it is mandatory or variable, is given an equal opportunity to be included in our sample. This approach provides a comprehensive and representative view of the variable code within the CSS projects. The exclusion of files without variable code is in line with systematic sampling, where irrelevant items are deliberately excluded to focus the analysis on the relevant aspects of the population. 

\bigskip

\noindent\textbf{Association Between Variable Code Information and Development History.} With information on the development history and variable code in our possession, we proceed on the task of associating developers with the variable code components. Our objective was to establish a dual association between developers: one that links them to the files they have changed, and another that connects them to the variable code they have impacted. This means that we can now identify developers who have actively added, modified, or removed variable code during the course of the CSS development. It is important to highlight that we do not differentiate the types of edits made to variable code. Although these edits may target distinct concerns—such as configuration directives, feature-related conditions, or general implementation logic—we adopt a unified view of variability-related contributions. This decision reflects a trade-off between granularity and robustness: differentiating edit types across systems with heterogeneous practices and tooling could introduce significant noise and bias, especially when feature representations are implicit or inconsistently maintained. Our unified treatment emphasizes overall maintenance activity related to variability and aligns with the source-level scope of this study. The concept of establishing associations between variables is fundamental in statistical analysis and research. It allows us to identify patterns and relationships within the data. By associating developers with variable code, we aim to understand the intricate dynamics of their involvement with variable code.

\subsection{Expertise Metrics}
\label{sec:expert_metrics}

To compare with the ground truth, we use two code expertise metrics: Degree of Authorship~\cite{fritz2014degree} and Ownership~\cite{bird2011don}. It is essential to acknowledge that the chosen expertise metrics, though not exhaustive, encompass fundamental concepts adopted within the field of software engineering~\cite{rahman2011ownership,bird2011don,fritz2014degree,avelino2019measuring, cury2022}.

\smallskip

\subsubsection{Degree of Authorship (DOA)} \label{ssb:doa}

To quantify developer expertise in the context of code authorship, we leverage the DOA metric proposed by \cite{fritz2014degree}. This metric computes the  score for a contributor (\textit{c}) within a specific file (\textit{f}) and software version. The equation to compute DOA is defined as follows:
\begin{equation}
DOA(\textit{c}, \textit{f}) = 3.293 + 1.098 \cdot FA + 0.164 \cdot DL - 0.321 \cdot \ln(1 + AC)
\end{equation}
where $FA$ (First Authorship) is a binary parameter indicating whether contributor \textit{c} created file \textit{f} (1 if true, 0 otherwise); $DL$ (Deliveries) is the number of code changes made by contributor \textit{c} in file \textit{f}; and $AC$ (Acceptances) is the number of code changes made by contributors other than \textit{c} in file \textit{f}.

The DOA metric considers three primary factors to assess a contributor's expertise level within a file. First, it acknowledges the significance of being the initial author of the file through the $FA$ parameter. Second, it accounts for the contributor's ongoing involvement by incorporating the number of code changes they made ($DL$). Finally, it balances these factors by considering the number of changes made by other contributors ($AC$) through a logarithmic term.

For the sake of simplicity, we computed and use hereafter in this work the normalized DOA ($DOA_N$) as given in~\cite{avelino2019measuring}:
\begin{equation}
DOA_N(c,f) = \frac{DOA(c,f)}{max(\{DOA(c',f) | c' \in changed(f)\})}
\end{equation}

In the $DOA_N$, \textit{changed(f)} denotes the contributors who created or edited a file \textit{f} up to a commit of interest. Therefore, $DOA_N \in [0..1]$, and values close to 1 are granted to the contributors with the highest absolute DOA among the contributors of a file. 
Based on the normalized DOA, the set of authors of a file \textit{f} is computed as given in~\cite{avelino2019measuring}:
\begin{align}
authors(f) \ = \ \{c \  & | \ c \ \in \ changed(f) \nonumber \\
                        & \land \ DOA_N(c,f) \ > \ 0.75 \nonumber \\
                        & \land \ DOA(c,f) \ \geq \ 3.293\}
\end{align}
The interpretation of DOA results depends on specific thresholds. This work uses 0.75 and 3.293 as thresholds to $DOA_N$ and DOA, respectively. In other words, a developer who achieves a value higher than the aforementioned is considered an author of the file. Otherwise, this developer is viewed as a contributor. We stem those thresholds from the work by~\cite{avelino2019measuring}.

\subsubsection{Ownership}

In addition to the DOA metric, another common approach to quantifying developer expertise within a file is Ownership. This metric, proposed by \cite{bird2011don}, computes the percentage of code within a file that a particular developer can be considered responsible for. The ownership metric is formulated as follows:
\begin{equation}
Ownership(c, f) = \frac{Commits_c(f)}{TotalCommits(f)} \times 100
\end{equation}
where $Ownership(c, f)$ is the Ownership percentage of developer $c$ in file $f$; $Commits_c(f)$ is the number of commits made by developer $c$ to file $f$; and $TotalCommits(f)$ is the total number of commits made to file $f$.

The Ownership metric offers a straightforward approach to assess developer expertise based on their contribution history within a file. The set of major contributors of a file \textit{f} is computed as:
\begin{align}
majors(f) \ = \ \{c \  & | \ c \ \in \ changed(f) \nonumber \\
                        & \land \ Ownership(c, f) \ > \ 5\%\}
\end{align}

A higher Ownership value indicates a greater level of involvement and potential expertise in a file. Similar to the DOA metric, interpretation of Ownership scores often relies on predefined thresholds. \cite{bird2011don} suggest a threshold of 5\%, where developers with ownership below 5\% are classified as minor contributors, while those exceeding 5\% are considered major contributors. %This threshold can be adjusted based on the specific context and project characteristics.

\smallskip

To extract the expertise metrics, we have crafted scripts that process all source code files, considering the entire development history of the subject systems. After processing the source code files, we generate a comprehensive dataset that contains expertise-related values for every developer within each source code file. This step aligns with established practices in quantitative research methods. The computation of expertise-related metrics DOA and Ownership is rooted in statistical methodologies that provide a quantitative assessment of developers' contributions and influence over specific source code components (i.e., files).

\subsection{Data Analysis Procedure} \label{sec:eval}

In this section, we delineate the data analysis procedures designed to effectively address the RQs of our study (see Section~\ref{sec:rq}). Our approach is grounded in robust statistical and methodological principles to ensure the validity and reliability of our findings.

\bigskip

\noindent\textbf{Developers Engagement in Mandatory and Variable Code (RQ1):} To comprehensively understand the work specialization dynamics among developers in relation to variable and mandatory code components, we implemented a systematic and longitudinal approach. We analyzed the full version history of each system to identify, for every commit of every developer, whether s/he contributed to variable or mandatory code. Based on this complete evolution data, we aggregated all contributions made by each developer throughout the observed timeframe. Only after this cumulative analysis were developers classified into one of three following categories, depending on the nature of their involvement:  

\begin{itemize}
    \item \textit{Generalists}: Developers exclusively engaged in modifying mandatory code throughout the entire development history.
    \item \textit{Specialists}: Developers exclusively dedicated to altering variable code components, considering the entirety of the development history.
    \item \textit{Mixed}: Developers who exhibited a profile by contributing to both variable and mandatory code segments (i.e., at least one commit in each type of code).    
\end{itemize}

We also examine whether the proportion of mandatory code relative to variable code helps explain the distribution of developer types. To do so, we computed correlation coefficients to assess linear associations and built regression models to evaluate the predictive power of the mandatory vs.\ variable code ratio on the presence of different developer categories. This approach allows us to quantify the extent to which structural characteristics of the codebase influence developer engagement patterns.

\bigskip

\noindent\textbf{Concentration of Workload on Variable Code (RQ2):} To address RQ2, we adopted a data aggregation approach informed by concentration statistics~\cite{gastwirth1972estimation}. Our primary goal was to analyze the equality or inequality in the distribution of work among developers in relation to variable code components.
In other words, the study investigates the concentration of variable code expertise among developers who actively engaged with variable code. This focus allows us to isolate the dynamics of expertise distribution within the subset of developers responsible for maintaining variability, which is critical for understanding bottlenecks and sustainability risks in CSSs. This analysis seeks to provide insights into the concentration of contributions, enabling us to make meaningful statements, such as \textit{"10\% of the specialist and mixed developers are responsible for over 70\% of the changes in variable code."}
For this purpose, we employed the Lorenz inequality, also known as the Lorenz curve, which is a well-established statistical method commonly used to examine income inequalities within a country's population~\cite{gastwirth1972estimation}. In our context, we repurposed this methodology to explore the concentration of changes in variable code per developer. For details of the Lorenz inequality, readers may refer to Lorenz's original work~\cite{gastwirth1972estimation}. 

To facilitate the comparison of Lorenz concentrations between different subject systems, we condensed this information into a single numerical metric known as the Gini coefficient ($g$)~\cite{dorfman1979formula}.
The Gini coefficient provides a scalar representation of the degree of distributional inequality concerning variable code changes among developers in a given system. It assumes values between 0 and 1, with $g=0$ signifying perfect equality, wherein a certain percentage of developers are collectively responsible for a proportionate amount of the variable code changes. Conversely, $g=1$ represents perfect inequality, indicating that a solitary developer shoulders the entire cumulative burden of variable code changes. By using the Gini coefficient, we can effectively compare different concentrations, and this metric encapsulates the concentration level in a single numerical value.

\bigskip

\noindent\textbf{Expertise Metrics and Variable Code Engagement (RQ3):}  To address RQ3, we employed a multi-step approach to examine whether developers designated as file experts by metrics like DOA and Ownership are involved at least once during the evolution history in modifying variable code within the files where their expertise is indicated. These metrics, detailed in Section~\ref{sec:expert_metrics}, were extracted for source files across the analyzed systems (described in Section~\ref{sec:systems}), forming the basis for our examination of the correlation between file expertise and variable code involvement. We emphasize that our goal in RQ3 is not to propose new expertise metrics, but to assess whether existing, general-purpose metrics—primarily file-based—remain effective in the face of variability. These results serve as a baseline for the development and validation of new variability-aware metrics in future work.

We compiled a list of developers who had contributed to variable code changes within each file over the system’s evolution. Then, we assessed whether developers identified as experts—those with high DOA or Ownership according to the thresholds defined in Section~\ref{sec:expert_metrics}—were included in this list of contributors working on variable code. To quantify the alignment between expertise metrics and actual developer contributions to variable code, we computed Precision and Recall~\cite{Ting2010}. Precision measured how many of the developers indicated by the metrics were indeed contributors to variable code, while Recall assessed how well the metrics identified all relevant contributors from the ground truth. These metrics enabled us to evaluate the reliability of expertise metrics in reflecting active developer engagement with variable code. As an external validity check, we cross-validate DOA and Ownership against six project‐level variables: (1) number of presence conditions, (2) mandatory‐code ratio, (3) variable‐code ratio, (4) generalist-developer percentage, (5) specialist-developer percentage, and (6) mixed-developer percentage. For each variable, we first apply the Shapiro–Wilk test to the 25 system values; if normality holds (p > 0.05), we use Pearson’s $r$, otherwise Spearman’s $\rho$, to correlate the variable with each expertise score. Through this analysis, we aimed to determine whether expertise metrics provide accurate recommendations or potentially miss critical developers contributing to variable code, offering insights into how these metrics can be refined for better alignment with practical development activities in CSSs. 

\bigskip

In addition, to investigate whether the variability intensity influences the concentration of developers' workload in variable code and the performance of expertise metrics, we partitioned the 25 systems into two groups to analyze the data from RQ2 and RQ3: 
\textit{high‐variability} (variable code > 40\% of total LOC) and \textit{low‐variability} (variable code < 10\% of total LOC). We set the \textit{high‐variability} threshold at > 40 \% and the \textit{low‐variability} threshold at < 10 \%, approximating the 75th ($\approx$ 37 \%) and 25th ($\approx$ 8 \%) percentiles, respectively, while ensuring a balanced group size (high: n=5; low: n=7). Subsequent analyses compare these groups in terms of precision, recall, and Gini coefficient.

Finally, it is important to notice that for RQ2 and RQ3, we restrict our analysis to the subset of developers who have contributed at least once to variable code. This includes developers classified as Specialists and Mixed, and excludes those who worked exclusively on mandatory code. This filtering ensures that our workload concentration and expertise evaluations are based only on relevant contributors to variability maintenance.

\section{Results} \label{sec:results}

This section presents the results to answer the RQs of our study. Section~\ref{sec:rq1} presents the results regarding the engagement of developers working on CSSs. We classify them as generalists, specialists, and mixed developers to analyze this division. Section~\ref{sec:rq2} describes the results of the concentration/distribution of work for different developers by using concentration statistics, allowing us to find imbalances. Finally, Section~\ref{sec:rq3} shows the results of the effectiveness of expertise metrics, by comparing file experts and variable code developers.

\subsection{Developers Engagement (RQ1)} \label{sec:rq1}

\begin{figure}[thb] 
\centering
    \includegraphics[width=\linewidth]{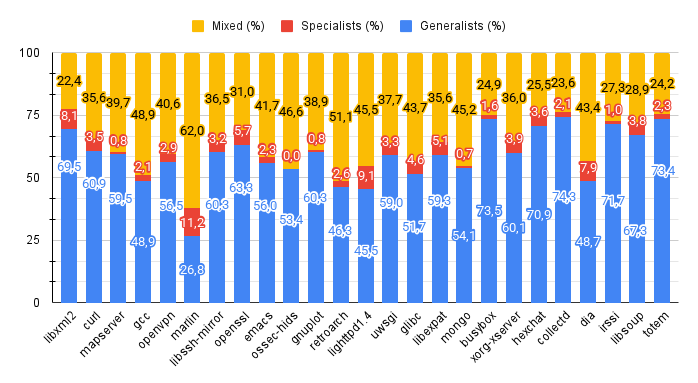} 
    \caption{Developer engagement across projects.}
    \label{fig:RQ1} 
\end{figure}

Figure~\ref{fig:RQ1} presents an overview of the total number of developers (\# Devs.) and the corresponding percentages of generalists, specialists, and mixed roles. With the results, we observe that 59\% of developers on average are classified as \textit{Generalists} across the subject systems. Their primary focus on mandatory code may limit their exposure to variable components, potentially impacting tasks such as maintenance and code reviews involving variable code.
\textit{Specialists} are no more than 4\% across subject systems, representing a minority group. This observation suggests that only a small fraction of developers specializes in variable code development.
\textit{Mixed Developers} account for approximately 37.5\% of developers on average. Those developers engage in both mandatory and variable code, providing a comprehensive perspective. However, their effectiveness can vary depending on their experience and exposure to variable code.

We further investigated whether the proportion of mandatory code relative to variable code ($M/V$) shown in Table~\ref{tab:project-stats} can explain the distribution of developer types by analyzing both correlations and regression models. The proportion of Generalists showed a strong positive Pearson correlation~\cite{freedman2007} with $M/V$ ($\rho = 0.68$, $p < 0.001$), while the proportion of Specialists was moderately and negatively correlated ($\rho = -0.54$, $p = 0.005$). Mixed developers also exhibited a negative correlation with $M/V$ ($\rho = -0.47$, $p = 0.018$), but this relationship is weaker.

To quantify these effects, we fitted three simple linear regressions using $M/V$ as the independent variable. The model predicting Generalist proportion yielded a positive slope ($\beta = +0.048$, SE = 0.007, $p < 0.001$), indicating that for each unit increase in $M/V$, the share of Generalists increases by 4.8 percentage points; this model explained 46\% of the variance ($R^2 = 0.46$). The model for Specialists returned a negative slope ($\beta = -0.019$, SE = 0.006, $p = 0.005$) with $R^2 = 0.29$, suggesting that higher $M/V$ ratios are associated with a moderate decline in Specialist presence. Finally, the regression for Mixed developers gave a slope of $\beta = -0.090$ (SE = 0.037, $p = 0.018$) with $R^2 = 0.22$, meaning that only 22\% of the variation in Mixed participation is explained by the structure of the codebase.

Taken together, these results support the intuitive expectation that more mandatory code is associated with more Generalists and fewer Specialists. However, the weaker statistical association for Mixed developers---both in correlation and explained variance---reveals a counter-intuitive pattern: the size of the mandatory or variable codebase are poor predictors of how many developers engage across both domains. This suggests that the presence and sufficiency of Mixed developers are shaped by additional factors beyond the proportion of code types.

Importantly, in preprocessor-based systems---the focus of this study---variable code is often intertwined with mandatory code through constructs such as \texttt{\#ifdef} blocks. In these cases, Mixed developers may be essential for reducing maintenance effort, as they possess at least a minimal understanding of the variable parts and are also likely familiar with the mandatory code that surrounds them. Their dual expertise may allow them to perform changes, review code and bug fixes more efficiently, especially in tangled feature locations where separation of concerns is limited.

The results of this study offer important insights for both research and practice in optimizing resource allocation and improving team collaboration in CSS projects. For the state of the art, these findings deepen our understanding of developers' engagement, particularly concerning how development teams handle variable and mandatory code. By identifying the different developer profiles and their engagement with code types, this study adds granularity to existing knowledge, aiding in the refinement of expertise metrics and team composition strategies. Researchers can use these insights to explore advanced methodologies for managing developer workloads, knowledge transfer, and resource allocation in software ecosystems.

For the state of the practice, the findings highlight the importance of strategically distributing developers across both code types to ensure better project outcomes and expertise management. Generalists, though dominant, may require additional training or support to effectively engage with variable code, while the limited number of specialists poses a risk of knowledge silos and potential bottlenecks. Identifying and addressing these gaps can lead to improved collaboration and reduced reliance on a small subset of developers. Additionally, mixed developers serve as a crucial bridge between mandatory and variable code tasks, and organizations should consider nurturing this role to enhance team flexibility and code maintainability.

%\medskip
\begin{tcolorbox}[left=0mm,right=0mm,boxrule=0.1mm,colback=gray!30!white]
\vspace{-0.1cm}
\textit{\textbf{Lesson Learned 1:}} Most developers focus on mandatory code, which may limit their expertise in variable code. This specialization gap underscores the need for better balance and training to prevent over-reliance on a few individuals and ensure effective handling of all code segments.
\vspace{-0.1cm}
\end{tcolorbox}

\medskip

\subsection{Concentration of Workload of Variable Code (RQ2)} \label{sec:rq2}

To answer RQ2, we explore the concentration and distribution of variable code development within the limited group of developers who work with variable code, namely mixed developers and specialists. We initiated our analysis by testing the normality of the distribution of variable code changes handled by individual developers using the Shapiro-Wilk test and an alpha-value of 0.05~\cite{SHAPIRO1965}. This test revealed that the data was probably not from a normal distribution. A non-normal distribution suggests a concentration of work, particularly if the distribution is skewed.

To visualize the distribution of work on variable code and assess its asymmetry, we employed violin plots, box plots, histograms, and Q-Q plots.\footnote{All plots are available as supplementary material in the replication package.} These plots, along with statistical measures of skewness and kurtosis, provided a more comprehensive assessment of the distribution and its skewness. This indicates that, on average, most developers interact with a relatively modest number of variable code elements, while only a select few developers deal with a more substantial portion. The analysis also demonstrated that the distribution of variable code is right-skewed, with the majority of developers engaging with a limited number of variable code elements, and a minority dealing with a more extensive portion. 

\begin{figure}[thb] 
\centering
    \includegraphics[width=\linewidth]{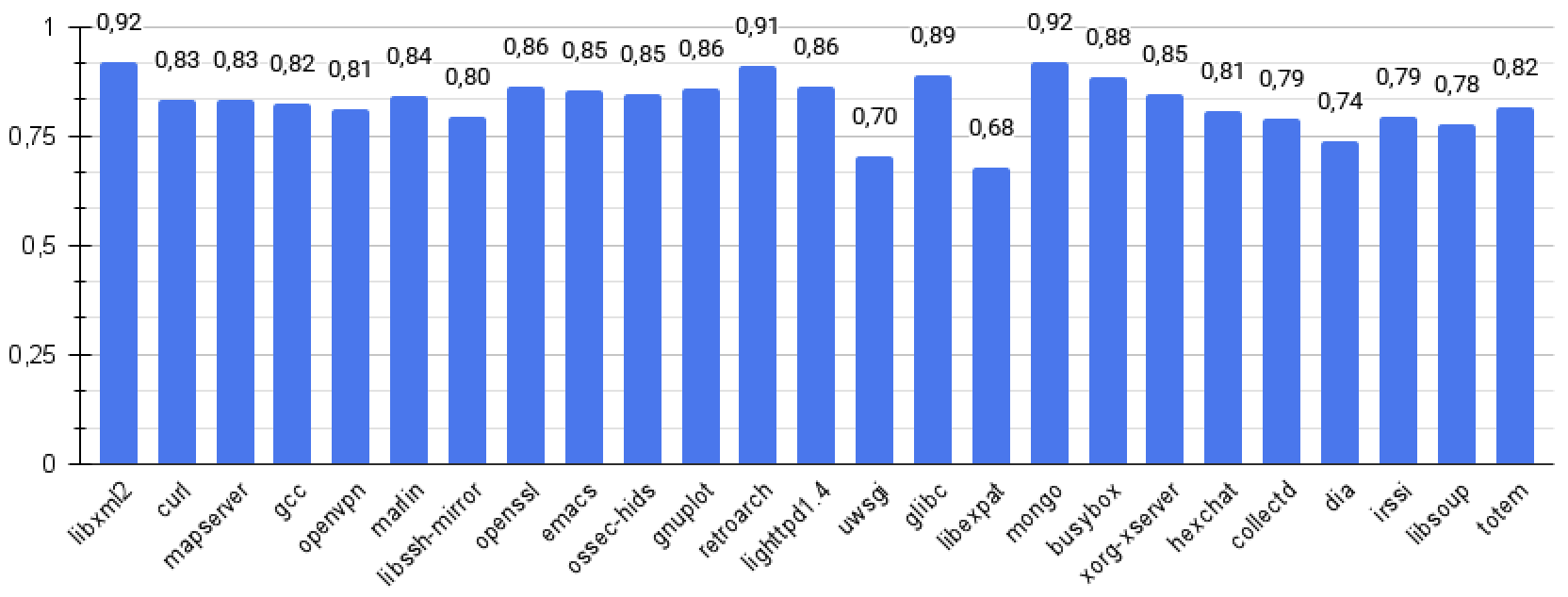} 
    \caption{Gini coefficient across projects.}
    \label{fig:RQ2} 
\end{figure}

With the goal of further investigating the concentration, we employed the Gini coefficient. This widely accepted measure of concentration offers a quantitative assessment of how variabilities are distributed among developers in the context of configurable systems. The Gini coefficient values are provided in Figure~\ref{fig:RQ2}. The average Gini coefficient across the subject systems is approximately 0.83. This result indicates that variable code development is not uniformly distributed among developers. Instead, it is highly concentrated in a subset of a limited number of developers who have knowledge about variable code.

Such high concentration can lead to several practical issues. For instance, if only a few developers handle the majority of variable code, it may result in bottlenecks or delays in development and maintenance tasks, as these individuals become overwhelmed with work. Additionally, it can create knowledge silos, where critical knowledge about variable code is confined to a small group, risking significant disruptions if these key developers leave or are unavailable. Moreover, this imbalance can affect team morale and productivity, as other developers may feel less engaged or perceive an inequitable distribution of responsibilities~\cite{Johnson:2021}.

\begin{figure}[htb]
\centering
    \includegraphics[width=0.7\linewidth]{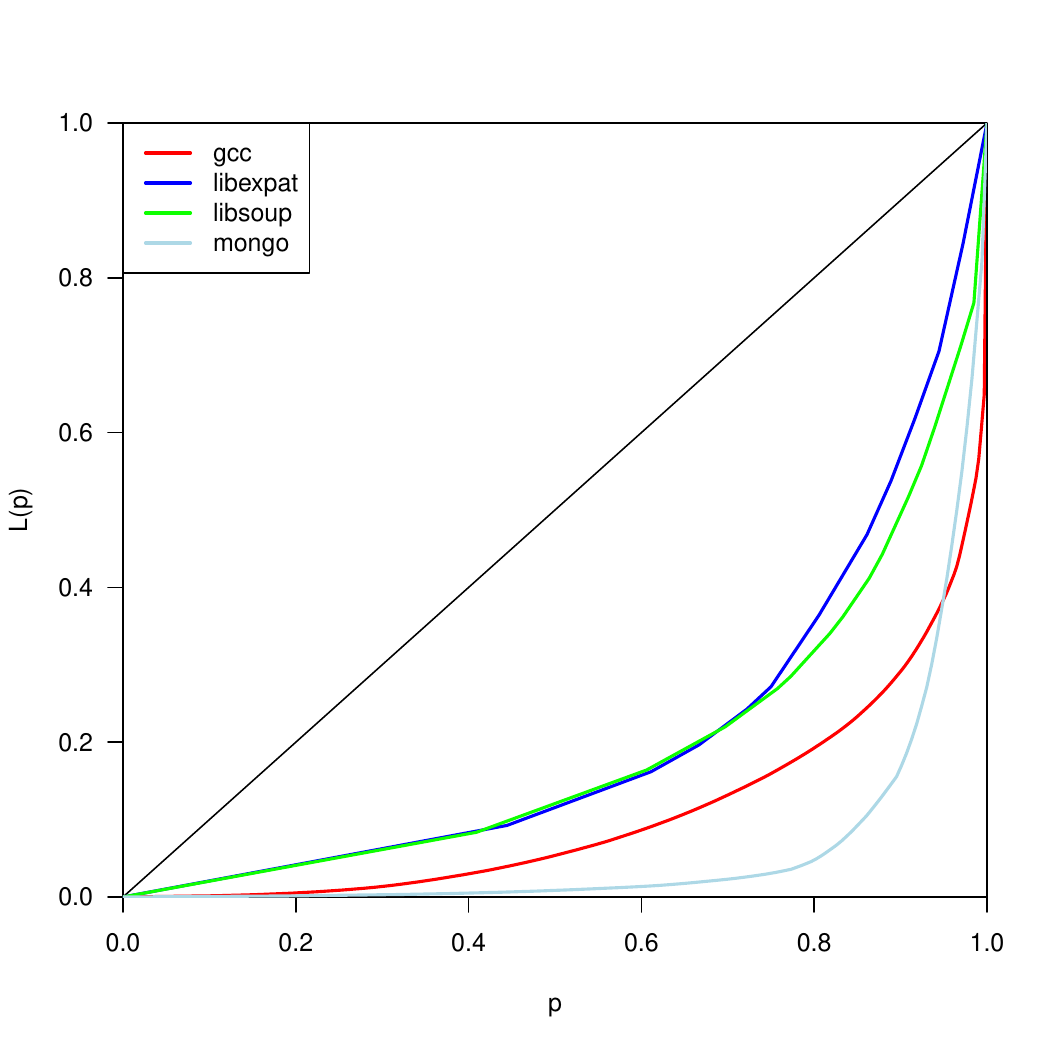} 
    \caption{Lorenz curve of four analyzed projects.}
     \label{fig:lcvariabilidadespordesenvolvedor}
\end{figure}

Figure~\ref{fig:lcvariabilidadespordesenvolvedor} presents the work concentration by plotting the Lorenz curve for four representative systems, selected based on (i) the groups defined in Section~\ref{sec:eval} (1 from upper quartile, 1 from lower quartile, 2 from interquartile), and (ii) number of configuration constants (the two systems with the highest and lowest numbers). The Lorenz curve is complementary to the Gini coefficient to represent inequality. The black line represents complete equality in the distribution of variable code to developers. The further the curve is from the line of equality, the more uneven the distribution. The Lorenz curves of the four systems corroborate with what was found in all subject systems and with the findings mentioned above.

The observed concentration of variable code development within a limited number of developers who actively engage with variability aspects presents a noteworthy consideration. By focusing on developers who engaged with variable code, we isolate the challenges of expertise silos specific to variability management. While generalists (aprox. 59\% of developers) avoid variable code entirely, their exclusion from RQ2 allows us to study the structural risks posed by those responsible for maintaining variability. This targeted analysis reveals that even within this subgroup, workload is highly concentrated, threatening project resilience.

In order to assess how variability intensity relates to workload concentration, we partitioned our subject systems into two groups based on the percentage of variable code (Section~\ref{sec:eval}): a high-variability group (projects with > 40 \% variable code: libxml2, curl, mapserver, gcc, openvpn) and a low-variability group (projects with < 10 \% variable code: xorg-xserver, hexchat, collectd, dia, irssi, libsoup, totem). We then compared the Gini coefficients for each group.

\begin{itemize}
    \item High-variability projects exhibit very high concentration of expertise, with Gini values ranging from 0.810 (openvpn) to 0.919 (libxml2). The average Gini for this group is approximately 0.84, indicating that a small subset of developers shoulders the vast majority of variable-code changes.

    \item Low-variability projects also display unequal distributions, but to a slightly lesser degree: Gini values range from 0.739 (dia) to 0.847 (xorg-xserver), with an average of about 0.80.
\end{itemize}

These results suggest a clear—but not overwhelming—trend: higher overall variability correlates with greater concentration of workload. In systems where variability comprises a larger fraction of the codebase, variable-code expertise becomes even more siloed among fewer individuals, exacerbating the risks of bottlenecks and single-point failures. Conversely, while low-variability projects still suffer from uneven distributions, the effect is somewhat attenuated, likely because the absolute volume of variable code is smaller and possibly more uniformly encountered by contributors.

While the general distribution of variable code development across different developer types might not seem concerning, the Gini coefficient analysis reveals a significant lack of uniformity in this distribution. This finding underscores several critical challenges. Developers with limited exposure to variable code may struggle with making informed decisions, identifying potential issues, and conducting thorough code reviews, all of which are crucial for maintaining code quality. Additionally, the concentration of responsibilities within a small group not only risks creating knowledge silos but also poses the threat of bottlenecks if these developers are unavailable. 

These results are consistent with existing literature on conventional software systems, which highlights concentration and workload imbalance among developers during the evolution history~\cite{Kuutila:2020}. However, the unique characteristics of CSSs necessitate further investigation to determine optimal distribution practices that address these challenges effectively.% within this context.

%\medskip
\begin{tcolorbox}[left=0mm,right=0mm,boxrule=0.1mm,colback=gray!30!white]
\vspace{-0.1cm}
\textit{\textbf{Lesson Learned 2:}} The concentration of variable code among a few developers aligns with issues found in conventional software systems, such as bottlenecks and knowledge silos. However, CSS projects may require tailored strategies for better distribution and management, specifically for variable code responsibilities.
\vspace{-0.1cm}
\end{tcolorbox}
%\medskip

\subsection{Association between file expertise and variable code (RQ3)} \label{sec:rq3}

Table~\ref{tab:performance-metrics} shows that the DOA metric achieves an average precision of 0.6, indicating that 60\% of developers recommended by this metric are likely to possess significant experience with variable code. This level of precision suggests that DOA is reasonably effective in identifying developers who have directly contributed to variable code. High precision is advantageous as it minimizes the inclusion of irrelevant developers, ensuring that recommended individuals are likely to have the requisite expertise. However, this strength also presents a limitation: a precision of 0.6 implies that 40\% of recommended developers might lack this expertise, potentially excluding some capable contributors. Ownership metrics exhibit a lower average precision of 0.5, indicating that only 50\% of the recommended developers are likely to have direct experience with variable code. This reduced precision suggests that Ownership may include a broader range of developers, some of whom may not have had interaction with variable code, considering the entire evolution history of the project. While this can be beneficial in capturing a wider pool of potential contributors, it also raises the risk of recommending developers who may not have the necessary depth of expertise.

\begin{table*}[t]
    \scriptsize
    \caption{Performance Metrics (Precision and Recall) for DOA and Ownership}
    \label{tab:performance-metrics}
    \centering
    \addtolength{\tabcolsep}{-2pt}
    \begin{tabular}{l
                    cc  % DOA
                    cc} % Ownership
      \toprule
      & \multicolumn{2}{c}{\textbf{DOA}} 
      & \multicolumn{2}{c}{\textbf{Ownership}} \\
      \cmidrule(lr){2-3} \cmidrule(lr){4-5}
      \textbf{Project} 
        & \textbf{Precision} & \textbf{Recall} 
        & \textbf{Precision} & \textbf{Recall} \\
      \midrule
      libxml2         & 0.67 & 0.24 & 0.45 & 0.41 \\
      curl            & 0.55 & 0.26 & 0.43 & 0.34 \\
      mapserver       & 0.64 & 0.41 & 0.73 & 0.75 \\
      gcc             & 0.53 & 0.62 & 0.52 & 0.75 \\
      openvpn         & 0.76 & 0.33 & 0.52 & 0.48 \\
      marlin          & 0.58 & 0.32 & 0.54 & 0.49 \\
      libssh-mirror   & 0.54 & 0.33 & 0.40 & 0.48 \\
      openssl         & 0.53 & 0.36 & 0.42 & 0.56 \\
      emacs           & 0.74 & 0.31 & 0.77 & 0.44 \\
      ossec-hids      & 0.76 & 0.37 & 0.42 & 0.63 \\
      gnuplot         & 0.57 & 0.24 & 0.56 & 0.40 \\
      retroarch       & 0.70 & 0.39 & 0.65 & 0.56 \\
      lighttpd1.4     & 0.75 & 0.35 & 0.64 & 0.41 \\
      uwsgi           & 0.65 & 0.24 & 0.42 & 0.40 \\
      glibc           & 0.62 & 0.56 & 0.52 & 0.78 \\
      libexpat        & 1.00 & 0.41 & 0.70 & 0.73 \\
      mongo           & 0.33 & 0.60 & 0.36 & 0.84 \\
      busybox         & 0.40 & 0.35 & 0.35 & 0.52 \\
      xorg-xserver    & 0.73 & 0.43 & 0.48 & 0.61 \\
      hexchat         & 0.57 & 0.21 & 0.21 & 0.29 \\
      collectd        & 0.33 & 0.38 & 0.30 & 0.56 \\
      dia             & 0.54 & 0.46 & 0.47 & 0.61 \\
      irssi           & 0.67 & 0.46 & 0.28 & 0.59 \\
      libsoup         & 0.43 & 0.58 & 0.33 & 0.79 \\
      totem           & 0.24 & 0.39 & 0.28 & 0.70 \\
      \midrule
      \textbf{Minimum}    & 0.24 & 0.21 & 0.21 & 0.29 \\
      \textbf{Maximum}    & 1.00 & 0.62 & 0.77 & 0.84 \\
      \textbf{Average}    & 0.59 & 0.38 & 0.47 & 0.56 \\
      \textbf{Median}     & 0.58 & 0.37 & 0.45 & 0.56 \\
      \textbf{Mode}       & 0.67 & 0.24 & 0.52 & 0.56 \\
      \textbf{Std.\ Dev.} & 0.17 & 0.11 & 0.15 & 0.15 \\
      \textbf{Skewness}   & --0.04 & 0.61 & 0.37 & 0.13 \\
      \textbf{Kurtosis}   & 0.69 & --0.18 & --0.44 & --0.96 \\
      \bottomrule
    \end{tabular}
\end{table*}

Recall further complicates the analysis. For DOA, a recall ranges from 0.2 to 0.6, with an average and median of 0.4. This recall level signifies that DOA may fail to identify up to 60\% of developers who have experience with variable code but are not highlighted as primary authors. The low recall indicates that DOA metrics might miss many developers who have relevant experience. Ownership metrics, on the other hand, exhibit higher recall, ranging from 0.3 to 0.8 with an average and median of 0.6. This higher recall suggests that Ownership is better at identifying a wider range of developers who have engaged with variable code. However, the broader scope might include individuals with no experience in variable code.

We partitioned our systems as described in Section~\ref{sec:eval} into a high-variability group (> 40 \% variable code: \textit{libxml2, curl, mapserver, gcc, openvpn}) and a low-variability group (< 10 \%: \textit{xorg-xserver, hexchat, collectd, dia, irssi, libsoup, totem}). For each group, we then computed the mean Precision and Recall for both the DOA and Ownership metrics.

High-variability systems exhibit substantially higher precision on both DOA (0.63 vs.\ 0.50) and Ownership (0.53 vs.\ 0.34), indicating that when the model predicts a developer as responsible, it is correct more often in these projects. However, recall is lower in the high-variability group for both DOA (0.37 vs.\ 0.42) and Ownership (0.55 vs.\ 0.59), suggesting that the model misses a larger fraction of true contributors in variability-dense codebases. In contrast, low-variability systems yield higher recall but at the cost of lower precision: the model casts a wider net—identifying more of the true DOA and Ownership instances—but also generates more false positives. This trade-off likely reflects the inherent structure of low-variability projects, where variability-related activities are fewer and more evenly distributed, making contributors easier to find but harder to pinpoint precisely.

This pattern suggests that when variable code comprises a large share of the codebase, expertise metrics become more “precise” (fewer false positives) but less “comprehensive” (more false negatives), potentially overlooking contributors to specialized or contextual code regions. Conversely, in low‐variability systems, metrics capture a broader set of contributors at the expense of precision. In high-variability systems, toolchains and task-assignment policies should emphasize expanding recall  (e.g., by incorporating additional heuristics) to avoid overlooking key variable-code contributors and probably reducing concentration workload. On the other hand, in low-variability systems, efforts might focus on increasing precision (e.g., by tightening threshold criteria or using more granular DOA and Ownership definitions) to reduce spurious attributions without sacrificing the decent recall already achieved.

\begin{figure}[htb]
\centering
    \includegraphics[width=0.7\linewidth]{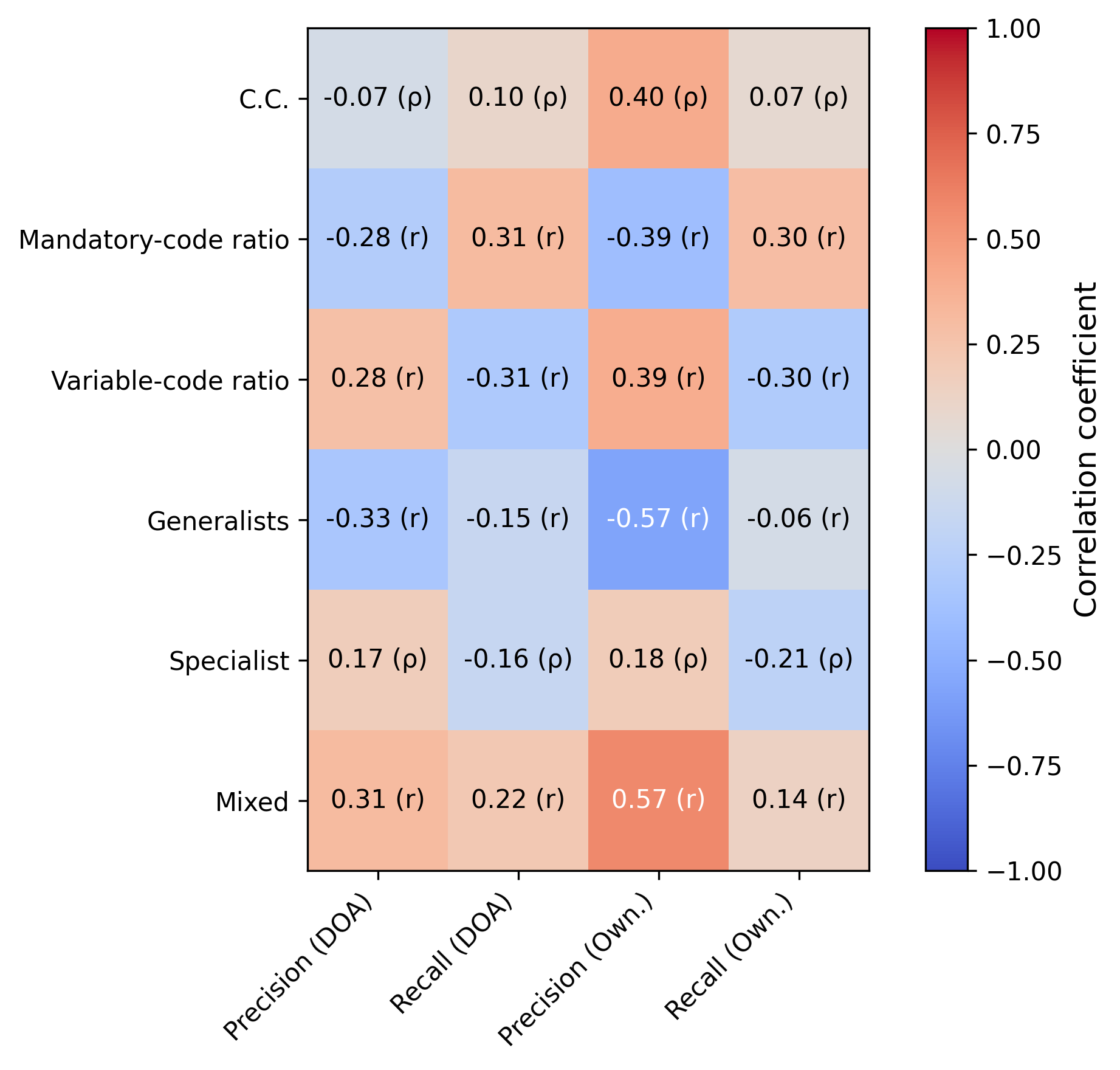} 
    \caption{Correlation of DOA and Ownership with project-level variables.}
     \label{fig:heatmap}
\end{figure}

To assess external validity, we next correlated our expertise metrics with six project‐level variables: (1) number of presence conditions, (2) mandatory‐code ratio, (3) variable‐code ratio, (4) generalist‐developer percentage, (5) specialist‐developer percentage, and (6) mixed‐developer percentage. We applied Shapiro–Wilk tests (N = 25) to select between Pearson’s $r$ (if $p>0.05$) and Spearman’s $\rho$ (otherwise). Figure~\ref{fig:heatmap} presents all coefficients.

None of the DOA metrics showed statistically significant correlations (all p > 0.05); the strongest non‐significant trends were a negative association of DOA precision with generalist‐developer percentage (r = –0.33, p = 0.10) and a positive association with mixed‐developer percentage (r = 0.31, p = 0.13). This suggests DOA captures fine‐grained, commit‐level behaviors—such as recency and edit granularity—that are orthogonal to coarse project characteristics and may benefit from incorporating variability‐aware weighting (e.g., presence‐condition density) to improve sensitivity in variability‐heavy modules. By contrast, Ownership precision correlates strongly with mixed‐developer percentage (r = 0.58, p = 0.003), generalist‐developer percentage (r = –0.57, p = 0.003), and number of presence conditions ($\rho$ = 0.40, p = 0.045), indicating that balanced expertise profiles, fewer generalists, and greater conditional complexity yield sharper ownership attribution.

Together, these findings indicate that these metrics capture different aspects of developer engagement beyond simple project‐level characteristics.

In summary, the trade-offs observed between precision and recall underscore a fundamental challenge in expertise identification within CSS projects. In the state of the art, these findings highlight the limitations of current metrics in capturing the full spectrum of developer expertise, thus achieving an optimal balance between accurately identifying experts and encompassing a broad range of contributors. This suggests a need for more refined metrics that can achieve better precision and recall, thereby providing a more accurate representation of developer expertise. In practice, the tendency for DOA to concentrate expertise recommendations on a smaller group of developers may lead to knowledge silos and potential bottlenecks, impacting overall project efficiency. Conversely, while Ownership’s higher recall captures a broader spectrum of contributors, it may result in less precise recommendations, affecting task allocation and productivity. Therefore, relying solely on these metrics may lead to an over-reliance on a small group of recognized contributors, risking bottlenecks and uneven distribution of variable code tasks. For CSS projects, there is a clear need for improved metrics or combined approaches that address these trade-offs, ensuring that expertise recommendations are both accurate and comprehensive.
\begin{tcolorbox}[left=0mm,right=0mm,boxrule=0.1mm,colback=gray!30!white]
\vspace{-0.1cm}
\textit{\textbf{Lessons Learned 3:}} Relying solely on existing expertise metrics can lead to incomplete or inaccurate identification of developer expertise. It is essential to refine these metrics or develop new ones to ensure a more precise and comprehensive assessment of expertise in CSS projects.
\vspace{-0.1cm}
\end{tcolorbox}

\section{Implications} \label{sec:implications}

The knowledge acquired by the analysis and results of our study can be directly translated to actionable implications to improve the practice of software engineering, discussed in what follows:

\bigskip

\noindent \textbf{Rethinking Developer Engagement Strategies.}
The disparity in how developers engage with variable versus mandatory code (RQ1) highlights an underlying challenge: the need for more sophisticated approaches to managing developer code engagement in CSS projects. Variable code is often more complex and volatile~\cite{ernst2002empirical}, requiring a more intentional effort to ensure adequate developer engagement. Our findings suggest that current practices may leave gaps, where variable code sections might be insufficiently maintained or understood by the broader team. In particular, we observed that while the proportion of variable code is moderately associated with the presence of Specialists, it has only a weak relationship with the proportion of Mixed developers, indicating that factors beyond code size drive the allocation of hybrid developer roles. This calls for a reevaluation of how developers are directed toward code areas based not only on workload or ownership, but on strategic importance to the system’s evolution. This is especially relevant in preprocessor-based systems, where variable code is often interleaved with mandatory code (e.g., via \texttt{\#ifdef} directives). In such cases, Mixed developers---those who contribute to both code types---can play a crucial role in reducing maintenance effort. Their dual familiarity allows them to understand and modify variable features in context, leveraging knowledge of the mandatory code that structurally and semantically surrounds them. This ability to bridge both domains can be instrumental in addressing tangled concerns and minimizing the risk of regressions in highly integrated systems. From a research perspective, this reveals an opportunity to investigate models of developer engagement that better account for the complexity and long-term maintainability of critical code areas such as those involving variability. In practice, teams may benefit from adopting adaptive engagement models, where developer assignments are dynamically adjusted based on evolving project needs, shifting the focus from static ownership to more flexible, system-wide stewardship models.

\bigskip

\noindent \textbf{Addressing the Structural Vulnerabilities of Expertise Silos.}
The concentration of variable code expertise among a small group of developers (RQ2) poses systemic risks not just to project timelines, but to the organizational resilience of software teams~\cite{avelino2016novel, Pfeiffer:2021}. This finding points to the existence of ``expertise silos'' that can create bottlenecks and heighten the project's vulnerability in case key developers leave or become unavailable~\cite{Motingoe2019}. While previous work in software engineering has addressed task allocation and knowledge sharing in general terms, our results call for a more nuanced understanding of how expertise silos emerge specifically around variable code. For teams and organizations, this underscores the need for actively dismantling such silos through structured knowledge dissemination practices. Beyond code reviews or mentoring, organizations should explore more formalized mechanisms such as rotating technical leadership or implementing practices that foster cross-functional knowledge acquisition. This has implications for software engineering research as well, suggesting a need for studies that investigate the lifecycle of expertise silos and the impact of their dissolution on team performance and CSS sustainability.

\bigskip

\noindent \textbf{Challenging Existing Expertise Metrics.}
Our analysis of expert developer engagement with variable code (RQ3) raises questions about the efficacy of existing expertise metrics. Current models often rely heavily on contributions to specific files, which may not reflect the real complexity or the types of work developers are engaging with over time in CSSs. The disconnect we found between ``designated'' experts and their actual engagement with variable code challenges assumptions about how expertise should be measured in CSS contexts. This has broader implications for both academia and industry, as it challenges the traditional reliance on file-level expertise metrics. Variabilities in CSS do not always align with the standard unit of abstraction used in programming languages, such as files or modules. Variable code often cuts across different files or components, making it difficult for current expertise metrics to capture the full scope of a developer's engagement with these areas. Expertise should therefore not be viewed as a static attribute tied to specific files, but as a dynamic quality that reflects a developer's ability to handle more nuanced and cross-cutting concerns, such as variabilities. There is a clear need to refine expertise models to account for these more intricate and overlapping abstractions, moving beyond file-level metrics to encompass broader areas of variability management and cross-component interactions. For researchers, this opens up a new line of inquiry into expertise models that consider variability and context as key factors in assessing developer engagement and effectiveness. In practice, development teams would benefit from adopting more granular and context-aware measures of expertise, incorporating elements such as the ability to solve problems across different layers of the system's variability. This would allow for better task allocation and responsibility assignment that accurately reflects the real challenges developers face in complex, variability-heavy codebases, ultimately improving system maintainability and reducing technical debt.

\bigskip

\noindent \textbf{Designing Tools to Support Context-Aware Task Allocation.}
One clear implication of our study is the inadequacy of current task allocation mechanisms adopted in the analyzed CSS projects, where the complexity of variable code and its disproportionate handling by certain developers may lead to inefficiencies (RQ2)~\cite{el2019metrics}. This finding motivates the development of new tools and approaches that can better account for the unique characteristics of CSS projects when assigning tasks~\cite{nundlall2021task}. These tools should not only distribute workload more evenly but should also incorporate context-sensitive factors, such as the complexity of the code being worked on and the frequency of changes in specific sections. From a research perspective, this opens the door to new algorithms for workload balancing that prioritize sustainability and resilience, in addition to immediate project efficiency. For practitioners, adopting tools that integrate real-time analysis of code complexity and developer engagement can ensure more informed decision-making in task assignment. This has the potential to reduce bottlenecks, improve team collaboration, and ultimately lead to better project outcomes.

\bigskip

\noindent \textbf{The Path Forward for Expertise and Engagement in CSS.}
Ultimately, the findings of this study highlight the need for a fundamental shift in how expertise and engagement are conceptualized in CSS projects. Traditional methods of assessing developer expertise—based on static metrics or historical contributions within a file—are no longer sufficient in the face of the dynamic and complex nature of variable code. Our study reveals that expertise must be seen as fluid, context-dependent, and constantly evolving. This realization calls for a deeper integration between expertise management and project management tools, allowing teams to monitor expertise in real-time and adjust accordingly. For the research community, this suggests an opportunity to reframe how we study developer expertise, shifting from static measures toward dynamic, context-aware models that account for the inherent variability in complex systems. For practitioners, embracing this new understanding means adopting continuous learning practices and fostering a team culture where expertise is not fixed but constantly developed and shared.

\section{Related Work} \label{sec:relatedwork}

%\noindent \textbf{C preprocessor.} 
The C preprocessor has been extensively criticized in the literature, with numerous studies discussing its detrimental effects on code quality and maintainability~\cite{adams2008aspect, ernst2002empirical, favre1995cpp, favre1997understanding, krone1994inference, spencer1992ifdef}. Efforts have been made to extract structural elements from source code, such as nesting, dependencies, and include hierarchies, and visualize them in dedicated views~\cite{krone1994inference, pearse1997experiences, spencer1992ifdef}. Researchers have also explored configuration views displaying selected portions of feature code to reduce complexity~\cite{atkins2002using, chu2003visual, hofer2010toolchain, kastner2008granularity, singh2007c}. Moreover, innovative approaches like color-coding to aid developers in handling variable code have been investigated~\cite{feigenspan2013background, rambally1986influence, oberg1992error, wuu1994merge}. 
Similar to previous work, our study aims to support the use of the C preprocessor. However, we take a different approach by focusing on leveraging workload and concentration distribution information in CSSs.

Several prior studies have investigated the impact of developer contributions to software quality. McDonald and Ackerman introduced the ``Line 10 Rule,'' a widely used heuristic for expertise recommendation~\cite{mcdonald2000expertise}. Other implementations such as Expertise Browser~\cite{mockus2002expertise} and Emergent Expertise Locator~\cite{minto2007recommending} offer alternatives to this rule. Our study employs the Degree-of-Authorship (DOA) metric~\cite{fritz2014degree} and Ownership~\cite{bird2011don} to identify experts, considering the code history. Rahman and Devanbu~\cite{rahman2011ownership} explored ownership and experience effects on quality in open-source projects, similar to our work but with a different operationalization of ownership and without considering preprocessor-based CSS. Likewise, Meneely and Williams~\cite{meneely2009secure} examined developer numbers' relationship to security vulnerabilities in the Linux kernel, while Avelino et al.~\cite{avelino2019measuring} analyzed code authorship in various systems, including CSS. However, neither study explored the suitability of expertise-related metrics for such systems. 
In addition, expertise metrics in software engineering have been studied extensively. Anvik et al. \cite{anvik06,anvik07} proposed techniques to identify developers capable of fixing specific bugs based on their bug-fixing history. Cury et al.\cite{cury2022} investigated methods to identify developers familiar with specific source code files.
Giordano et al. \cite{giordano2024adoption} investigate how code reuse mechanisms, such as inheritance and delegation, are adopted over time. The study identifies recurring usage patterns and confirms that these mechanisms have a significant impact on software quality and maintainability.
However, these techniques are primarily tailored for traditional codebases and may not effectively capture expertise in variable code.

Configurable Software Systems (CSSs) add another layer of complexity to the software development process. Several works have examined the challenges posed by CSSs, particularly in managing variability, improving maintainability, and ensuring quality. For example, Kästner et al. \cite{kastner2008granularity} and \cite{apel2013book}  studied feature-oriented programming and variability-aware analysis in CSSs. Their work highlights the unique challenges of handling feature interactions and the preprocessor’s impact on variability.

Further, Liebig et al. \cite{liebig2010analysis} examined the prevalence of conditional compilation and its implications for developers, emphasizing the difficulties of reasoning about feature interactions. Similarly, \cite{thum2014classification} proposed techniques to verify feature model correctness, a key aspect of CSS quality. Passos et al. Passos et al. \cite{passos2016coevolution} analyzed the evolutionary patterns of CSSs, showing how variability impacts software evolution and developer expertise over time.

The work of Kröher et al.~\cite{KROHER2023111737} investigates the intensity and frequency of variability-related changes during the evolution of Software Product Lines (SPLs). By analyzing four KConfig-based SPLs, including the Linux kernel, the authors found that variability changes—such as those in variability models, source code, and build files—occur infrequently and typically affect only small portions of the artifacts. The results indicate that variability evolves in a stable manner, suggesting that verification tools and processes can be optimized to handle smaller and less frequent modifications. This study contributes to the understanding of SPL maintenance by highlighting stability patterns in variability evolution.

Comparing with our prior paper~\cite{Milano2024}, which provided only an aggregate view of workload distribution in CSS, this work (1) introduces the Gini coefficient for formal inequality analysis, (2) performs per‐project concentration analyses to reveal system‐specific patterns, (3) deepens mixed‐developer quantification by measuring variable‐code commit shares, (4) applies the Shapiro–Wilk test with Q–Q and violin plots for robust statistical validation, (5) enhances cross‐project comparisons of both concentration and expertise metrics, and (6) derives actionable recommendations for tool support—such as workload imbalance alerts and dual‐domain developer features—that were absent in the earlier study.

Our study builds on this growing body of literature by focusing specifically on the suitability of expertise-related metrics in CSSs, which have been largely underexplored. Unlike previous studies, we emphasize the sociotechnical aspects of CSSs, leveraging the DOA metric to capture expertise distribution and workload concentration. This approach provides new insights into how expertise can be effectively measured and utilized in systems where variability is a central concern.

By integrating insights from works on developer expertise, variability management, and CSS quality, our research contributes to a deeper understanding of expertise in CSSs and underscores the need for more specialized metrics tailored to their unique characteristics.

\section{Threats to Validity} \label{sec:threats}

\bigskip
\noindent\textbf{Conclusion Validity} (treatment-outcome relationship) concerns violated statistical test assumptions. We mitigated this by employing non-parametric tests with minimal assumptions about data distribution. Sampling bias due to commit selection criteria was addressed by clearly defining inclusion/exclusion criteria, ensuring transparency and alignment with best practices.

\bigskip
\noindent\textbf{Internal Validity} (causal effect of independent variables) is threatened by historical events, where past reviewers might have concentrated knowledge. We mitigated this by analyzing systems with many developers and variabilities. Script and tool limitations, where errors in custom data extraction scripts or the modified preprocessor could occur, were addressed through testing and validation.

\bigskip
\noindent\textbf{Construct Validity} (generalizability across constructs) is limited by our focus on C language source code. While this limits generalizability to other languages, C is a prevalent language with many preprocessor-based systems on GitHub. Additionally, inadequate pre-operational explication of constructs, where variable and mandatory code definitions are unclear, could threaten validity.  We mitigated this by providing clear and detailed construct definitions aligned with existing research.

\bigskip
\noindent\textbf{External Validity} (generalizability to broader populations) is limited by the use of open-source systems, which might differ from proprietary systems. However, these open-source systems are widely adopted and relevant. Additionally, limited generalizability to a broader range of software systems is a potential threat. We addressed this by providing comprehensive information about the subject systems and their contexts, while also using real-world, open-source configurable systems from GitHub with established research backgrounds.
\section{Conclusions}
\label{sec:conclusion}

This paper presented an in-depth investigation into the role of developer expertise in managing variable code CSSs. Through an empirical analysis of 25 CSS projects, comprising over 450,000 commits and nearly 10,000 developers, we uncovered critical insights about the distribution of variable code, the concentration of expertise, and the limitations of current expertise metrics. Our findings reveal a significant imbalance in how variable code is handled: 83\% of variable code is maintained by only a few developers, with 59\% of developers never interacting with variable code throughout the project's history. This disproportionate concentration of expertise creates a bottleneck, where a small subset of developers bears the responsibility for a crucial yet complex portion of the codebase. Such imbalances can pose risks for long-term maintainability and scalability, as the system becomes increasingly dependent on the availability and capacity of these key individuals. Furthermore, we demonstrated that existing expertise metrics, which primarily rely on file-level contributions, struggle to accurately reflect the expertise required to manage variable code. With only 50\% to 60\% of precision and 40\% to 60\% of recall in identifying developers who engage with variable code, these metrics fail to capture the intricacies of developer interaction with the scattered and conditionally included code segments that define CSSs. This highlights a fundamental mismatch between the static, file-based nature of current expertise metrics and the dynamic, interleaved structure of variable code in CSSs.
From a practical standpoint, our findings suggest that refining expertise metrics to consider alternative units of abstraction could enable more effective task assignment and a more equitable distribution of workload among developers, thereby reducing the risks associated with expertise concentration. Although this study primarily employs a quantitative approach, we recognize that qualitative methods—such as manual code reviews or developer interviews—could offer deeper insights into why certain developers specialize in variable code. However, conducting interviews or case-study analyses would entail a fundamentally different methodology and resource commitment, which lies beyond the scope of this work. Future research should explore these qualitative dimension, alongside the development of dynamic expertise models and alternative methods for measuring engagement with variable code to further address the issues identified in this study. Moreover, longitudinal analyses leveraging the full commit history could reveal how individual developers’ involvement with variable versus mandatory code changes over the course of their careers—e.g., whether early specialization in one domain predicts later role transitions—thereby enriching our understanding of expertise evolution in configurable systems. Finally, future work should cross-validate alternative expertise models—such as developer network centrality, code review participation, or feature-model contributions—and compare them to the results presented in this paper. These efforts will be critical for advancing our understanding of expertise distribution and ensuring the long-term sustainability of CSS projects.

In summary, this research highlights the limitations of existing expertise metrics in the context of CSSs and points to the need for new approaches that can better capture the complexity and variability of software systems. By refining these metrics and fostering a broader distribution of expertise, both academia and industry can address the challenges posed by variable code, improving the overall sustainability and evolution of CSS projects. Future research should explore the development of dynamic expertise models and alternative methods for measuring engagement with variable code to further address the issues identified in this study.

\section*{Data-availability Statement}

A reproduction package~\cite{replicationpackage} for the study, including the implementations of the scripts for data extraction, the raw data, graphics, and the experimental results, among others, is available on Zenodo.

\section*{Acknowledgement}

Bruno B. P. Cafeo acknowledges the support received from FAEPEX (Fundo de Apoio ao Ensino, à Pesquisa e Extensão) under Grant No. 3404/23 and 2382/24. 

\section*{Declaration of generative AI and AI-assisted technologies in the writing process}

During the preparation of this work the author(s) used OpenAI's ChatGPT (GPT-4 version; available at \url{https://chat.openai.com/}) in order to draft, refine language, and organize our narrative structure. After using this tool/service, the authors reviewed and edited the content as needed and take full responsibility for the content of the published article.

\bibliographystyle{elsarticle-num-names}
\bibliography{references}

\end{document}